\documentclass[12pt]{article}
\usepackage{authblk}

\usepackage{graphicx}
\usepackage{a4wide}
\usepackage{palatino}
\usepackage{amsmath}
\usepackage{amssymb}
\usepackage{aas_macros}
\usepackage{cite}

\def\ll#1#2{\tilde{\lambda}_{#1}.\tilde{\lambda}_{#2}}
\def\llss#1#2{\tilde{\lambda}_{#1}.\tilde{\lambda}_{#2}\,\boldsymbol{\sigma}_{#1}\boldsymbol{\sigma}_{#2}}
\def\vec#1{\boldsymbol{#1}}
\def\dd{\mathrm{d}}
\counterwithout{figure}{section}

\begin{document}
\title{Resonances in the quark model%
\footnote{To appear in {\it Few-Body Systems}, special issue 
	on {\it Critical Stability}}
}
\author{Jean-Marc Richard}
\affil{Institut de Physique des 2 Infinis de Lyon,
Universit{\'e} Claude Bernard (Lyon 1), CNRS-IN2P3\\
4 rue Enrico Fermi, 69622 Villeurbanne, France}
\author{Alfredo Valcarce}
\affil{Departamento de F{\'i}sica Fundamental,
 Universidad de Salamanca\authorcr
{\small\sl E-37008 Salamanca, Spain}}

\author{ Javier Vijande}
\affil{Unidad Mixta de Investigacin en Radiof{\'i}sica e Instrumentaci{\'o}n Nuclear en Medicina\authorcr{\small\sl (IRIMED)
Instituto de Investigaci{\'o}n Sanitaria La Fe (IIS-La Fe)}\authorcr
{\small\sl Universitat de Valencia (UV) and IFIC (UV-CSIC) Valencia, Spain}}
\date{\small \today}
\maketitle
\begin{abstract}
 A discussion is presented of the estimates of the  energy and width of resonances in constituent models, with focus on the tetraquark states containing heavy quarks.
\end{abstract}
%


\section{Introduction}
There are many candidates for multiquark exotics or crypto-exotics.\footnote{The cryto-exotic states have quantum numbers accessible to quark-antiquark or three-quark configurations, but  differ from ordinary hadrons}\@
For a review, see, e.g., \cite{Lebed:2016hpi,Richard:2016eis,Liu:2019zoy,Brambilla:2019esw}. The  $T_{cc}^+$ with double charm, the $XYZ$ states with hidden-charm and the heavy pentaquarks have motivated several studies. For the two latter categories, there are resonances lying above their lightest threshold consisting of a charmonium and a light hadron, and their study requires a treatment that is suited for states in the continuum.

In the literature, there are many approaches to the physics of hadrons: quark model, lattice QCD, sum rules, molecules, etc.  It is challenging to study whether the quark model, if treated carefully, can provide a unified picture of ordinary hadrons and resonances.

Several methods have been proposed to estimate the properties of resonances: direct search in the complex-energy plane, analytic continuation from bound states obtained by artificially increasing the attraction or imposing an external confinement $W$, etc. For a real confinement, i.e., $v(r)\to v(r)+g\,W(r)$ for some pairs, where $v(r)$ is the interparticle potential, a suitable analytic continuation is performed  to $g\to 0$. A popular alternative consists of applying an absorbing confinement $v(r)\to v(r)-i\,\eta\,W(r)$, see, e.g., \cite{1993JPhB...26.4503R,PhysRevA.72.052704}. Then one may look either at $\eta\to0$ or  values of the energies that are stationary with respect to variations of $\eta$, as recommended, e.g., in \cite{Masui:2002zd}.
The choice of the absorbing potential $\eta\,W(r)$ to
 minimize reflections has been discussed in the literature, see, for example, 
 \cite{Manopoulos2002} for a one-dimensional case within the Jeffreys-Wentzel-Kramers-Brillouin approximation. However, the study leads to a singular potential that is not easily handled in numerical applications

So far, the applications to the quark model are mostly based on extensions of the variational methods designed for bound states. In particular, if
\begin{equation}\label{eq:trial}
 \Psi(\vec x, \ldots)=\sum_{k=1}^N \gamma_k \,\exp(-a_{k,lj}\,r_{lj}^n/2)~,
\end{equation}
is a trial function for bound states, the stabilization method (or real scaling) looks at plateaus when the range parameters are rescaled as
\begin{equation}
 \label{eq:real-sc}
 a_{k,lj}\to\alpha\, a_{k,lj}~,
\end{equation}
 and the minimal distance between two levels with avoided crossing provides a rough estimate of the width~\cite{1981JChPh..75.2465S}. The complex scaling method, in its variational variant,  corresponds to a rotation $a_{k,lj}\to \exp(-n\,i\,t)\,a_{k,lj}$, and the search for real and imaginary parts of the energy that are stationary when the rotation angle $t$ is varied. Complex scaling applied to a resonance of energy $E=E_{\rm r}-i\,\Gamma/2$ requires
 an angle $t\ge -\arg(E)/2$ to make the wave function square integrable. $t$ should be
 large enough to include the resonance, but not too large to avoid rapid oscillations of the asymptotes. In the variational
 variant, one merely searches whether there is a range of $t$ that
 corresponds to a plateau for $E_{\rm r}$ and $\Gamma$ \cite{Wang:2022yes},
 and if it exists, one can verify that it satisfies the above condition on $t$. Of course, a combination of real and complex scaling, namely
 \begin{equation}\label{eq:real-plus-complex}
  a_{k,lj}\to \alpha\,\exp(-n\,i\,t)\,a_{k,lj}~,
 \end{equation}
can be implemented, with both $\alpha$ and $t$ tuned to stabilize best  the resonance. In \eqref{eq:trial}, $\vec x, \ldots$ denote the Jacobi variables describing the relative motion, the  $r_{lj}=|\vec r_j-\vec r_l|$ are  the interparticle distances, $n=1$ is often used for small Coulomb systems, and $n=2$ is preferred for larger systems or in presence of confinement.

In \eqref{eq:real-sc} and \eqref{eq:real-plus-complex}, the real scaling parameter $\alpha$ depends on the choice of the starting parameters $a_{k,lj}$, which are tuned to describe the bound-state spectrum, if any, or are adjusted empirically. Such ``secret of recipe'' is not always clearly revealed in the literature dealing with resonances.
\section{Benchmark calculations of resonances}
\subsection{Bardsley potential}
\begin{figure}[ht!]
 \centerline{%
 \includegraphics[width=.3\textwidth]{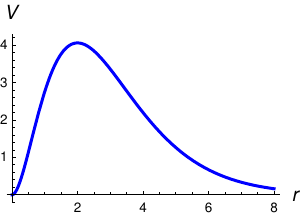}\quad
  \includegraphics[width=.3\textwidth]{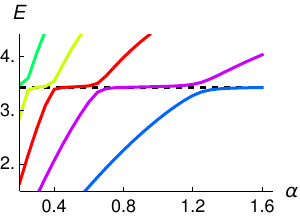}\quad \includegraphics[width=.35\textwidth]{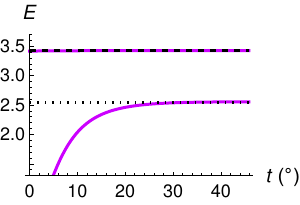}}
 \caption{Left:Bardsley potential. Center: Real scaling, energies as a function of $\alpha$. Right: real scaling ($\alpha=0.8$) supplemented by a complex rotation of angle $t$, showing the real part of the energy and the imaginary part  multiplied by $-200$.  The dotted lines are the exact values.}
 \label{fig:Bardsley1}
\end{figure}

It reads \cite{1974JPhB....7.2189B}
\begin{equation}\label{eq:Bradsley}
 H=\frac{\vec p^2}{2}+V(r)~,\qquad V(r)=V_0\,r^2\,\exp(-r)~,
\end{equation}
with most numerical results given for $V_0=7.5$, which is shown in Fig.~\ref{fig:Bardsley1}.  A detailed study is done, e.g., in \cite{2013JPhA...46h5303K} and is repeated here with another choice of basis.
In Fig.~\ref{fig:Bardsley1} are shown the results obtained with $N=8$ Gaussians, i.e., $n=2$,  in \eqref{eq:trial}. They agree perfectly with  \cite{2013JPhA...46h5303K}.

\begin{figure}[ht!]
 \centerline{
 \includegraphics[width=.35\textwidth,height=.25\textwidth]{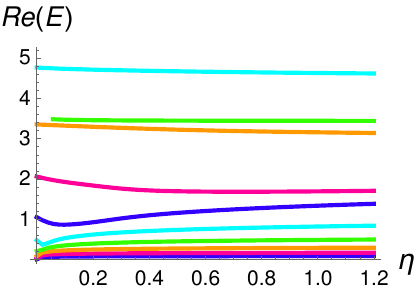}\qquad
 \includegraphics[width=.35\textwidth,height=.25\textwidth]{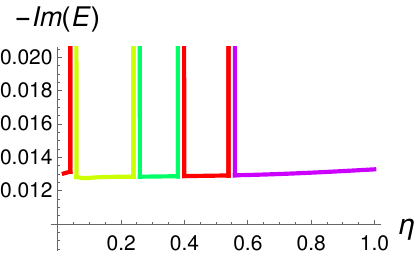}}
 \caption{Real (left) and imaginary (right) parts of the levels of the Bardsley potential supplemented by an absorptive confinement $-i\,\eta\,(r--r_c)^2\,\Theta(r-r_c)$ with $r_c=5$.}
 \label{fig:Brad-abs}
\end{figure}

The resonances of the Bardsley potential can also be studied with a complex confinement \cite{1993JPhB...26.4503R}, namely as the limit of the spectrum of
\begin{equation}
 H \ \to \ H_\eta=H - i\,\eta\,W(r)~,
\end{equation}
with $\eta$ real being the strength of a real confining potential such as $W=r^2\,\Theta(r-r_c)$ or $W=(r-r_c)\,\Theta(r-r_c)$, the radius $r_c$ being slightly outside the peak of the main potential $V(r)$. If $H_\eta$ is solved by numerical integration, one gets the results shown in Fig.~\ref{fig:Brad-abs}.
There is a dramatic stability of one of the levels, or more precisely, of an energy that is endorsed by successive levels. This is a kind of ``level rearrangement'' in the language of exotic atoms~\cite{2007IJMPB..21.3765C}.

\begin{figure}[ht!]
 \centerline{
 \includegraphics[width=.35\textwidth,height=.25\textwidth]{Fig5.pdf}\qquad
 \includegraphics[width=.35\textwidth,height=.25\textwidth]{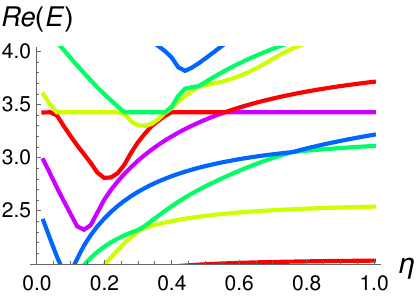}}
 \caption{Same a Fig.~\ref{fig:Brad-abs}, with a Gaussian expansion. }
 \label{fig:Brad-abs-gem}
\end{figure}

Interestingly, the method of absorbing confinement also works if associated with the Gaussian Expansion Method (GEM). It can be checked that a small basis, if suitably adapted, already provides good results. The ones presented in Fig.~\ref{fig:Brad-abs-gem} correspond to the comfortable choice of a large basis of $N=20$ terms with the range parameters $a_k$ in \eqref{eq:trial} ranging from $a_1=0.01$ to $a_N=100$ in a geometric series. As noted by Masui and Ho \cite{Masui:2002zd}, when the absorbing potential is used in connection with GEM or similar expansion on a finite basis, the best results are not obtained as $\eta\to0$ but rather as values at stationary points.

\subsection{Hazi-Taylor potential}
\begin{figure}[ht!]
 \centerline{
 \includegraphics[width=.4\textwidth]{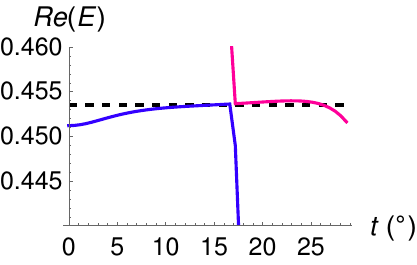}
 \hspace*{.2cm}
  \includegraphics[width=.4\textwidth]{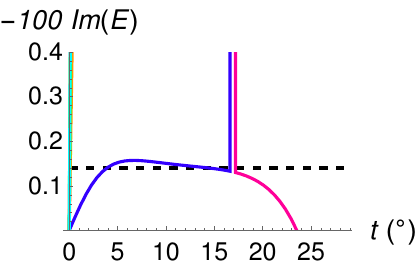}
  }
 \caption{Real and imaginary part of the energy as a function of the rotation angle $t$ for the potential of Hazi and Taylor of Eq.~\eqref{eq:HT} with $\lambda=0.19$. The overall real scaling is $\alpha=0.55$.  The dotted lines correspond to the exact values as given in \cite{1970PhRvA...1.1109H}.}
 \label{fig:HaziT}
\end{figure}

A similar model was proposed by Hazi and Taylor \cite{1970PhRvA...1.1109H}, namely
\begin{equation}\label{eq:HT}
 V(x)=\begin{cases}
        \, x^2 & \text{if} \quad  x<0\\
        \,x^2\,\exp(-\lambda \,x^2) & \text{if}\quad  x\ge 0~.
     \end{cases}
\end{equation}
We tested the method of real-plus-complex scaling by reproducing their results in the case $\lambda=0.19$. See Fig.~\ref{fig:HaziT}.
We used a basis of Gaussians
\begin{equation}
 \label{eq:basisHT}
 u(x)=\sum_{i=1}^N(\gamma_i\,+ x\,\delta_i)\,\exp(-a_i\,x^2)/2)~,
\end{equation}
with the even and odd sectors coupled by the potential term, and the range parameters $a_i$  in a geometric series. With only $N=5$ terms in the expansion, one gets near $\alpha=0.55$ and $t=15^\circ$ a stationary point, clearly indicating the existence of a resonance with the right energy.

\subsection{Remarks about the harmonic oscillator basis}
\begin{figure}
 \centerline{\raisebox{.23cm}{\includegraphics[width=.4\textwidth]{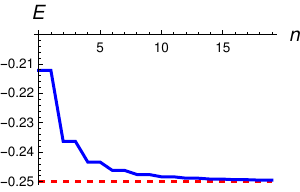}}
 	\qquad\qquad\includegraphics[width=.4\textwidth]{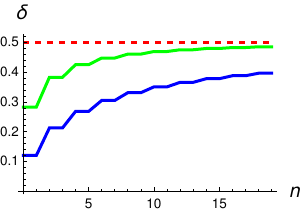}}
 \caption{Variational estimate of the ground state of the positronium using a basis of HO states, as a function of the number of terms. Left: energy, right: wave function at the origin, either directly from the variational wave function (blue) or using the Schwinger formula (green). The red dashed line indicates the exact value. \label{fig:CbHO}}
\end{figure}

Note that in the benchmark cases presented in the literature, it is often made use of an expansion on a harmonic oscillator basis (HO). It was also adopted in early quark-model calculations. However, it was noticed that the variational wave function based on HO hardly converges towards the exact ones both at very large and very short distance. The latter is needed to estimate the hyperfine corrections and some decay widths. Nowadays, the Gaussian expansion method is widely adopted, which does not suffer from this problem.  To illustrate this difficulty in the HO expansion, let us solve the Coulomb problem for two particles of mass $m=1$ and charge $e=\pm 1$. The exact  solution for the ground state is $E_{\rm ex}=-1/4$ for the energy and $u_{\rm ex}(r)=r\,\exp(-r/2)/\sqrt2$ for the  reduced radial function. If one uses a HO expansion
\begin{equation}\label{eq:var-ho}
 u_n(r)=\sum_{i=0}^n \gamma_i\,v_i(a,r)~,\qquad
 v_i(a,r)\propto \exp(-a\,r^2/2)\,\mathcal{H}_{2\,n+1}(r\,\sqrt{a})~
\end{equation}
where the normalized HO states  $v_i$ involve Hermite polynomials $\mathcal{H}$. As seen in Fig.~\ref{fig:CbHO}, there is a reasonable convergence of the energy towards the exact value. On the other hand, the short-range correlation, $\delta=u'(0)^2$, if read directly from \eqref{eq:var-ho}, hardly converges to the exact value $\delta_{\rm ex}=1/2$. Of course, if one uses the RHS of the Schwinger identity~\cite{Quigg:1979vr}
\begin{equation}\label{eq:Schw}
u'(0)^2=\int_0^\infty u(r)^2\,V'(r)\,\dd r~
\end{equation}
which expresses $\delta$ as the expectation value of the derivative of the potential $V(r)$, one gets a much better approximation.

A related side remark is the observation that the variational calculation progresses stepwise, namely, $E_{2\,N+1}=E_{2\,N}$ for the energy and similarly for the other quantities. This was already pointed out in the literature \cite{Moshinsky:348628,1976JPhG....2..357M}. In short, the stationarity condition $\partial \langle\Psi(a, \vec x, \ldots)|H|\Psi(a, \vec x, \ldots)|\rangle/\partial a$ implies that $H$ does not couple $\Psi$ to its derivative with respect to the range parameter $a$, and, in the case of an expansion on a HO basis, this implies a vanishing coupling of $\Psi_{2\,N}$ and $\Psi_{2\,N+1}$. The staircase character is clearly seen in Fig.~\ref{fig:CbHO}.

\subsection{Resonance in \boldmath{$\mathrm{H}^-$}\unboldmath}
In the community of ``critical stability'', the negatively charged hydrogen ion $\mathrm{H}^-$ is the first milestone, as the stability cannot be established using a naive factorized wave function $\Psi=f(r_1)\,f(r_2)$. Historically, the stability was demonstrated either by breaking and restoring the symmetry, namely $\Psi=f(r_1)\,g(r_2)+g(r_1)\,f(r_2)$ or by introducing an explicit anticorrelation, i.e., $\Psi=f(r_1)\,f(r_2)\, h(r_{12})$. See, e.g., the reviews \cite{1957qmot.book.....B,2010AmJPh..78...86H} for the abundant literature, in particular the contributions by Chandrasekhar and Hylleraas.

Of course, the precision is improved by combining the breaking-plus-restoration of symmetry and the account for anticorrelation, and by adding such terms, i.e.,
\begin{equation}
 \Psi(\vec r_1,\vec r_2)=\sum_{k,\ell}\gamma_{k,\ell}\,\left\{\exp[-(a_k\,r_1+b_k\,r_2+c_\ell\,r_{12})/2]+1\leftrightarrow 2\right\}~.
\end{equation}
It can readily be seen that with just a few terms, one reproduces the ground-state energy. For instance, with  5 pairs of $\{ a_k,\, b_k\}$ and 2 values of $c_k$, the best variational energy is $E_{\rm var}= -0.527727$ to be compared to the exact
$E_{\rm ex}=-0.527751\dots$, in atomic units~(a.u.).

\begin{figure}[ht!]
 \centering
 \includegraphics[width=.32\textwidth]{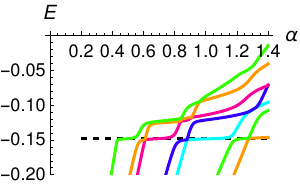}
\
 \includegraphics[width=.32\textwidth]{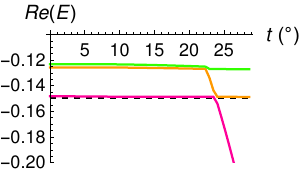}
 \
 \raisebox{-.25cm}{\includegraphics[width=.32\textwidth]{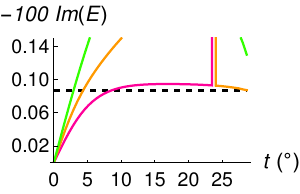}}
 %
 \caption{Left: estimate of the resonance of H$^-$ near $E=-0.15$ by real scaling. Center and right: complex scaling with the range coefficients multiplied by $\alpha\,\exp(-i\,t)$. Here $\alpha=0.8$. The dotted line corresponds to the most sophisticated estimates in the literature.\label{fig:Hminus}}
\end{figure}

The spectrum of H$^-$ includes a resonance with  energy near $E=-0.15$.
Starting from the above wave function,  the real-plus-complex scaling can be applied (here $n=1$), with the results shown in Fig.~\ref{fig:Hminus}.
They are very close with the numbers given by very sophisticated methods \cite{2008PhRvA..77d4502V,1974PhRvA..10..729B}. The real scaling clearly demonstrates the existence of a resonance near $E=-0.15$. The complex rotation of the range parameters confirms this resonance and gives a good estimate of its width. Note the avoided crossings both for real scaling and complex rotation, again reminiscent of the level rearrangement of levels in exotic atoms when the strength of the short-range potential is varied \cite{2007IJMPB..21.3765C}.
\subsection{Resonance in the Helium atom}
As studied, e.g., in \cite{2018JPhB...51x5001S}, there is a resonance of the Helium atom with a complex energy $-0.79484\ldots -i\,0.001245\ldots$
For this $S$-wave resonance of He, we have adopted a slight variant of the wave-function, namely
\begin{equation}
 \Psi=\sum_{i,\ell} \gamma_{i,\ell} \, r_{12}^\ell\,\left[\exp(-a_i\,r_1-b_i\,r_2)+\exp(-b_i\,r_1-a_i\,r_2))\right]~.
\end{equation}
In a first step, we have adjusted the range parameters to reproduce the ground state. With 4 pairs of $\{a_,b_i\}$ and $\ell=0,\,1$ or $2$, one obtains $E_0=-2.90368$ very close to the benchmark value $E_0=-2.903722\dots$  by Kinoshota, as cited by Bethe and Salpeter~\cite{bethe2013quantum}.

\begin{figure}[ht!]
 \centerline{
 \includegraphics[width=.32\textwidth]{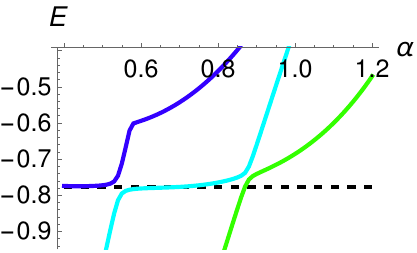}
 \
 \includegraphics[width=.32\textwidth]{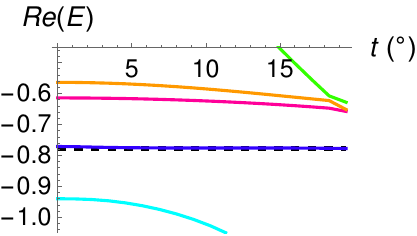}
 \
 \includegraphics[width=.32\textwidth]{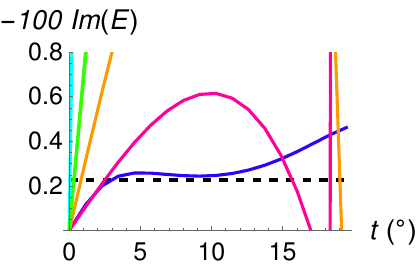}}
 \caption{Real and complex (with $\alpha=0.5$) scaling scan of the Helium resonance near $E=-0.778$ (in a.u.). The dotted lines correspond to accurate results in the literature~\cite{2011PhRvA..84c2515C}.}
 \label{fig:He-res}
\end{figure}

This wave function is taken as the starting point for the study of the resonance at $E=-0.079484\dots -I \, 0.001245\dots$, see, e.g., \cite{2011PhRvA..84c2515C}.
Then we apply a combination of real and complex scaling. The results are shown in Fig.~\ref{fig:He-res}, and compared to the literature.

\begin{figure}[th!]
 \centerline{
 \includegraphics[width=.35\textwidth]{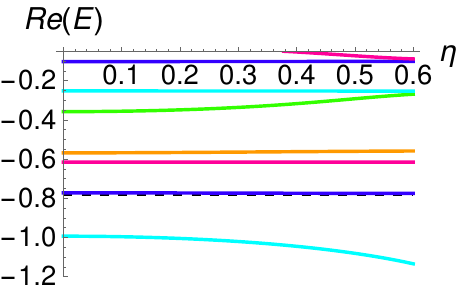}\hskip 1cm
  \includegraphics[width=.35\textwidth]{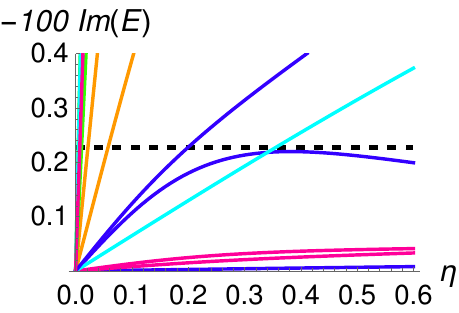}
  }
 \caption{Complex absorption applied to the Helium resonance near $E=-0.778$ in a.u.  Left: real part. Right: imaginary part.}
 \label{fig:Helium-w}
\end{figure}

This Helium resonance is also seen in the method of complex absorption. For instance, one can add to the pairs with Coulomb attraction an imaginary confinement
\begin{equation}\label{eq:quad-conf}
 W=-i\,\eta\,\sum_{j=1}^2 \Theta(r_j-r_c)\,(r_j-r_c)^2~.
\end{equation}
With the wave function determined in the real case ($\eta=0$), one gets the results shown in Fig.~\ref{fig:Helium-w}, corresponding to $r_c=10$. When $\eta$ is varied, the real part of the resonance is very stable, and the imaginary part exhibits a broad minimum near the exact value.

The conclusion of this section is that resonances can be calculated using a well-chosen basis of reasonable size combined to a simultaneous use of stabilization and complex rotation, or complex absorption.
\section{Resonances in the quark model}
There is an abundant literature on multiquark hadrons, in particular dealing with the $XY\!ZT$ states discovered during the last decades. While the  quark model is acknowledged as the simplest and most efficient picture of deeply bound states such as $bb\bar u\bar d$,  a ``molecular approach'' is
more often advocated for resonances. A tentative unification implies a good picture of the ordinary hadrons constituting the threshold and their long-range interaction.

A group in Japan~\cite{Hiyama:2005cf,Hiyama:2018ukv} has applied the method of real scaling (or stabilization) using a standard potential model
\begin{equation}
 \label{eq:potq}
 V=-\sum_{i<j} \left[\ll{i}{j}\, v(r_{ij}) + \llss{i}{j}\,\frac{v_{ss}(r_{i,j})}{m_i\,m_j}\right]~,
\end{equation}
that contains a spin-independent interaction $v$ and a spin-spin one, $v_{ss}$, where typically $v(r)$ has a Coulomb and a linear term, and $v_{ss}(r)$ is a smeared contact interaction.

The crudest modeling focused on the chromomagnetic interaction in \eqref{eq:potq}. Assuming that the kinetic energy $k_i$  and the chromoelectric interaction result in a kind of effective mass $\tilde  m_i$,  SU(3) symmetry for the light and strange constituents, and assuming further that the short-range correlation factor $\langle v_{ss}\rangle$ is the same in $H$ than in ordinary mesons or baryons, Jaffe diagonalized the spin-color operator
\begin{equation}
\sum_{i<j} \llss{i}{j}~,
\end{equation}
and made the following observations
\begin{itemize}
 \item For light scalars, the exotic states with isospin $I=2$ are pushed up in the spectrum, and thus are presumably very broad. This makes plausible the description of some of the lightest scalars as multiquarks \cite{Jaffe:1976ig}.
 \item For the 6-quark state $uuddss$ with spin-parity $0^+$, the chromomagnetic attraction is larger than twice that of spin-1/2 baryon \cite{Jaffe:1976yi}. A state $H=uuddss$. This exotic~$H$ was searched for in many experiments and further studied in a flurry of papers~\cite{Carames:2013hla,Nakazawa:2023xad}.
\end{itemize}

A detailed survey of the mechanisms pushing down the $H$ about 150\,MeV below its lowest baryon-baryon threshold shows that the $H$ does not survive  a realistic breaking of SU(3) and  a self-consistent treatment of the short-range correlations in each hadron coming from a rigorous treatment of the spin-independent part of the Hamiltonian. See, e.g., \cite{Oka:1983ku,Rosner:1985yh,Karl:1987cg}. Nevertheless, the last word has perhaps not been said, yet. Some recent lattice calculations find for the $H$ a mass close to the $\Lambda\Lambda$ threshold, and motivate further investigations, see, e.g., \cite{PadmanathMadanagopalan:2021exb} and refs.~there.

Shortly after the first speculations based on chromomagnetism, it was realized that the chromoelectric interaction can also produce some bound multiquarks. In a configuration $QQ\bar q\bar q$ with constituent masses $M$ for $Q$ and $m$ for $q$, the ground state becomes stable, i.e.,
\begin{equation}
 QQ\bar q\bar q < Q\bar q+Q\bar q~,
\end{equation}
if the mass ratio $M/m$ is large enough \cite{Ader:1981db}. The critical $M/m$ is very large, but if one combines chromoelectric and chromomagnetic effects, it is found that $cc\bar u \bar d$ is at the edge of stability \cite{Carlson:1987hh,Janc:2004qn,Vijande:2009kj}. And the state $T_{cc}^+$ with minimal quark content $cc\bar u\bar d$ was found by the LHCb collaboration~\cite{LHCb:2021vvq}, 40 years after the first prediction!

The improvement of stability when $M/m$ increases in the chromoelectric model
\begin{equation}
 H=\frac{\vec{p}_1^2+\vec{p}_2^2}{2\,M}+\frac{\vec{p}_3^2+\vec{p}_4^2}{2\,m}-\frac{3}{16}\sum_{i<j} \ll{i}{j}\,v(r_{ij})~
 \label{eq:chromoelectricH}
\end{equation}
is shared with hydrogen-like molecules if the potential energy in \eqref{eq:chromoelectricH} is replaced by
\begin{equation}\label{eq:potC}
 V=\sum _{i<j} \frac{q_i\,q_j}{r_{ij}}~.
\end{equation}
In atomic physics, any equal mass configuration $\mu^+\mu^+\mu^-\mu^-$ is stable by a small margin, but $M^+M^+m^-m^-$ becomes more stable is the mass ratio $M/m$ departs from 1. Indeed, one can decompose the Hamiltonian into a $C$-even and a $C$-odd part
\begin{equation}
 H=\left[\sum_i\frac{ \vec p_i^2}{2\,\mu}+V\right]+\left(\frac{1}{4\,M}-\frac{1}{4\,m}\right)\left(\vec p_1^2+\vec p_2^2-\vec p_3^2
-\vec p_4^2\right)=H_{\rm even}+H_{\rm odd}~,
\end{equation}
where $2\,\mu^{-1}=M^{-1}+m^{-1}$. The full Hamiltonian $H$ and its even part $H_{\rm even}$ have the \emph{same threshold}. Thus the odd term lower the ground state of $H$ and improves the binding, whenever $V$ is the chromoelectric interaction of the quark model or the Coulomb interaction \eqref{eq:potC}.

Starting from the equal mass case, one can arrange the  breaking of other symmetries, for instance of particle identity as $\mu^+\mu^+\mu^-\mu^-\to M^+m^+M^-m^-$. But then, the lowering of the energy of $H$ due to the odd-term also decreases the energy of the $M^\pm m^\mp$ atoms and actually the threshold benefits more from the effect, and the molecule becomes less stable and eventually unstable as $M/m$ increases. In atomic physics, the stability is lost at $M/m\simeq 2.2$ \cite{1997PhRvA..55..200B,2005PhR...413....1A}.
Note  that before becoming unstable, $M^+m^+M^-m^-$ becomes Borromean, as near $M/m=2$, both $M^\pm m ^\pm m^\mp$ and $M^\pm m ^\pm M^\mp$ ions are not bound~\cite{2003PhRvA..67c4702R}. Here, ``Borromean'' does not mean that all subsystems are unbound, but that one cannot build the molecule by adding the constituents one by one with stable intermediate states.

Once the stability of, say, the ground state $bb\bar u\bar d$, one may wonder whether there are some excitations in the continuum. A related investigation is whether there are resonances for configuration without bound state, such as $cc\bar c\bar c$, that are continuously suggested from an experimental point of
view 
\cite{CMS:2023owd}.

\begin{figure}[ht!]
 \centering
 \includegraphics[width=.5\textwidth]{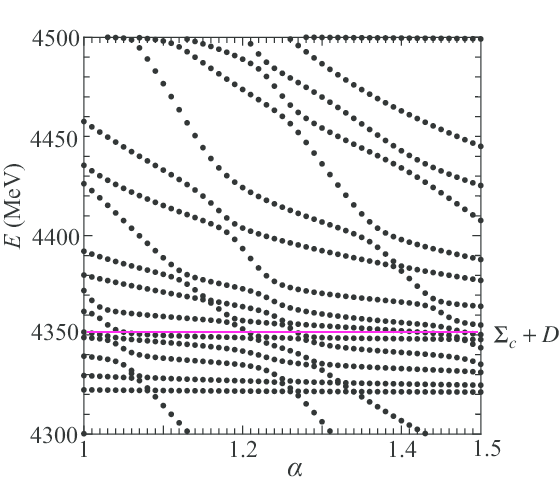}
 \caption{Stabilization graph for a  $c\bar c qqq$ configuration, from Ref.~\cite{Hiyama:2018ukv}}.
 \label{fig:Hiyama-penta}
\end{figure}

Hiyama et al.\ have calculated the spectrum of some pentaquark states using a potential of type \eqref{eq:potq}. See \cite{Hiyama:2005cf,Hiyama:2018ukv}. They used
real scaling, and obtained clear plateaus in the stabilization graphs. See Fig.~\ref{fig:Hiyama-penta}.
This calculation clearly demonstrates the possibility of separating the genuine resonances from the states that just mimic the continuum, with, however, two restrictions. First, the spectrum such as Fig.~\ref{fig:Hiyama-penta} involves a  very large basis of Gaussian states. Also the width is estimated from a formula by Simons \cite{1981JChPh..75.2465S} that relates this width to the minimal distance of levels at avoided crossing, and is not very accurate.

The method of real scaling has been applied to other configurations, for instance \cite{Meng:2020knc,Meng:2023for}. Alternatively, complex scaling was used in \cite{Lin:2023ihj,Wang:2022yes,2023PhRvD.108g1501W,Meng:2024yhu}.

In \cite{Wang:2022yes}, the S-wave spectrum of $cc\bar c\bar c$ is studied, using a potential by Barnes et al.~\cite{Barnes:2005pb}. Resonances are found in the $J^P=0^+,\,1^+$ and $2^+$ sectors.   For $J^P=2^+$ two states  are found, with $M-i\,\Gamma/2= 7068.5-i\,41.8$ and  $7281.3 -i\,45.6\,$MeV. 
Similar results, although with some additional resonances, have been obtained in Ref.~\cite{Wu:2024euj} using a Gaussian expansion method with complex scaling.

\begin{figure}[ht!]
 \centerline{
 \includegraphics[width=.32\textwidth]{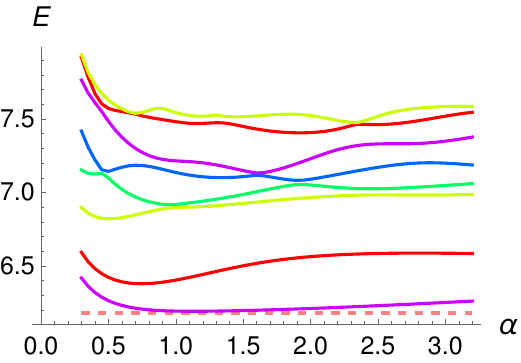}
\
 \includegraphics[width=.32\textwidth]{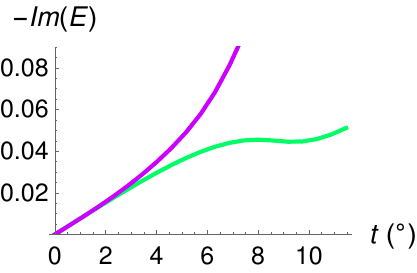}
 \
  \includegraphics[width=.32\textwidth]{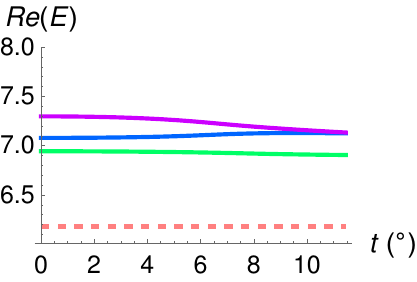}
  }
 \caption{Study of $cc\bar c\bar c$ resonances with the BGS potential~\cite{Barnes:2005pb}, using a Gaussian expansion of the wave function. Left: real scaling, center and right: real and imaginary energy using complex scaling with $n=2$ and $\alpha=1.55$ in \eqref{eq:real-plus-complex}, atop a first empirical selection of the range parameters.}
 \label{fig:BGS-2plus}
\end{figure}

An analysis using a small set of Gaussians confirms, indeed, the presence of a resonance near $6.9\,-i\,0.04\,$GeV, as seen in Fig.~\ref{fig:BGS-2plus}, which corresponds to a real scaling by $\alpha=1.55$ before applying the complex rotation. Note that at this point of our study, and survey of the literature, the choice of the parameters before real and complex scaling remains rather empirical. Either one uses a very large number of Gaussians, as to fully cover the range of interest, or one minimizes the ground-state as to approach the threshold. See, e.g., the discussion by Varga {\sl et al.} \cite{2008PhRvA..77d4502V} in another context.

\begin{figure}[ht!]
 \centerline{

 \includegraphics[width=.32\textwidth]{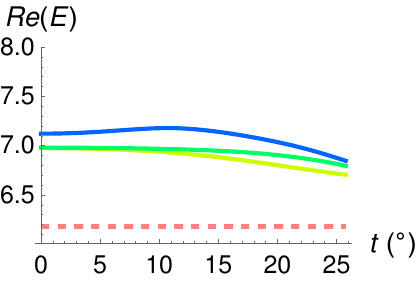}
 \
  \includegraphics[width=.32\textwidth]{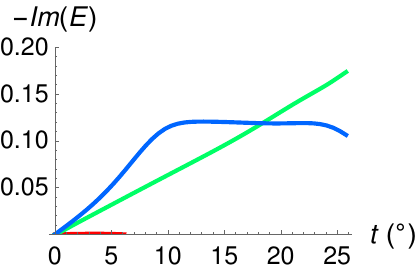}
  }
 \caption{Same as Fig.~\ref{fig:BGS-2plus}, but for $\alpha=2.55$.}
 \label{fig:BGS-2plusp}
\end{figure}

Then for $\alpha=2.55$, the complex energy $E$ shows the pattern of Fig.~\ref{fig:BGS-2plusp}. Both real and imaginary parts are stationary, indicating a resonance near $E=7.1-i\,0.12\,$GeV. The width is, however, somewhat larger than the one found in \cite{Wang:2022yes}. A larger basis is perhaps necessary to get a better estimate. More details will be given in a forthcoming article~\cite{RVV:2024}.

Another representation of the results is shown in Fig.~\ref{fig:BGS-2plusp}, with the trajectories of some levels in the complex energy plane.
\begin{figure}[ht!]
 \centering
 \includegraphics[width=.45\textwidth]{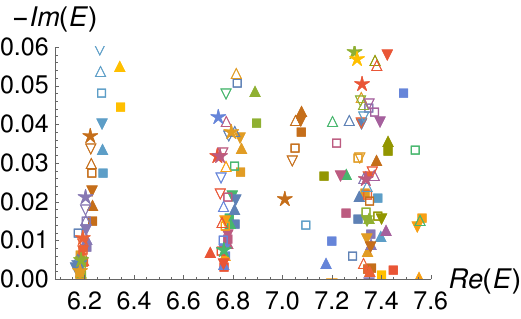}
 %
 \caption{Complex scaling scan of the $2^+$ sector of $cc\bar c\bar c$ with the BGS potential. The symbols  $\{\blacksquare ,\blacktriangle ,\blacktriangledown ,\square ,\triangle ,\triangledown ,\star \}$ correspond to increasing values of the rotation angle $t$ from 8 to 20$^\circ$, and the color to the various levels in this energy range.}
 \label{fig:BGS-2pluspp}
\end{figure}
We confirm an accumulation of points near $E=7.0-I\,0.04\,$GeV, as noted by Meng et al.~\cite{Meng:2023for}.

\begin{figure}[ht!]
 \centerline{
 \includegraphics[width=.35\textwidth]{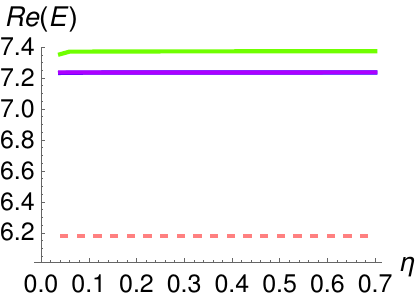}
 \qquad
  \includegraphics[width=.35\textwidth]{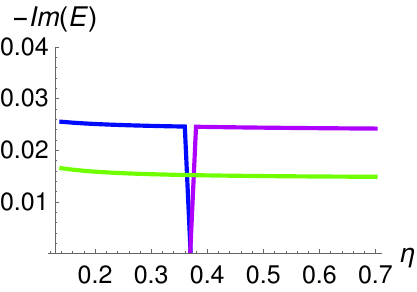}}
 \caption{Selected levels of the spin 2 $cc\bar c\bar c$ configuration with the BGS potential, as a function of the strength $\eta$ of the external absorbing confinement.}
 \label{fig:BGS-spîn2-w}
\end{figure}

The occurrence of resonances in this complex-energy range is also confirmed by the method of complex absorption. An example is shown in Fig.~\ref{fig:BGS-spîn2-w}. We adopted an absorbing confinement similar to \eqref{eq:quad-conf}, with $r_c=4\,\mathrm{GeV}^{-1}$ applied to all pairs in the color $\bar33$ channel.

Note the rearrangement observed in the imaginary part of the energy when the strength $\eta$ is varied. The real parts of the rearranging levels are indistinguishable.

\section{Concluding remarks}
As for the technical aspects, our aim was to provide some handy tools to estimate quark-model resonances. For benchmark potentials such a the Bardsley potential, it is shown that a Gaussian expansion with just a few term allows one to reproduce the most sophisticated results of the literature. A combination of real and complex scalings gives access to the width of the resonance. The method of complex absorption works also very well.

For atomic systems such a H$^-$ or He,  a combination of real and complex scaling turns out very efficient, and the results are nicely confirmed by the method of complex absorption, though it requires some tuning. One can note that a reasonable accuracy is already obtained with just a few terms in the variational wave function.

For multiquark systems, the situation is more delicate. Already demonstrating the existence of a second bound state (as predicted in some models of $bb\bar u\bar d$) is a difficult task. And it is even harder to get access to the resonances of $QQ\bar Q\bar Q$ with $Q=b$ or $c$. It is embarrassing to confess that a basis of more than a thousand terms is necessary. It is however interesting to observe that the method of real-plus-complex scaling, in connection with a Gaussian expansion, and the method of absorbing confinement lead to similar results. A more thorough comparison of the methods will be given elsewhere~\cite{RVV:2024}.


As for the quark dynamics, the question is whether long-range and short-range effects can be unambiguously combined within a single model. The quark potentials such as \eqref{eq:potq} accounts rather well for the quarkonium spectrum and properties. Their ability to describe the interaction among two quarkonia remains uncertain.

\subsection*{Acknowledgements}
The authors acknowledge enlightening correspondence with Prof.~M.~Oka.
This work has been partially funded
by the Spanish Ministerio de Ciencia e Innovaci\'on (MICINN) and the
European Regional Development Fund (ERDF) under 
contracts PID2019-105439GB-C22, PID2022-141910NB-I00,
and RED2022-134204-E.

%

\end{document}